\documentclass{tADP2eArXiv}

\usepackage{epstopdf}% To incorporate .eps illustrations using PDFLaTeX, etc.
\usepackage{subfigure}% Support for small, `sub' figures and tables
\usepackage{url} % Included by the authors to allow special characters such as _, & in urls.
\usepackage{bbm}
\usepackage{color}
\usepackage{wrapfig}
\usepackage[colorlinks=true,citecolor=red]{hyperref}

\newcommand{\be}{\begin{equation}}
\newcommand{\ee}{\end{equation}}
\newcommand{\ben}{\begin{eqnarray}}
\newcommand{\een}{\end{eqnarray}}
\newcommand{\bes}{\begin{subequations}}
\newcommand{\ees}{\end{subequations}}
\newcommand{\bF}{\begin{figure}}
\newcommand{\eF}{\end{figure}}

\let\baraccent=\= % rename builtin command \= to \baraccent
\def \cv{\mathrm{Cov}}

\def\tr#1{{\rm{Tr}}\left[#1\right]}

\theoremstyle{plain}
\newtheorem{theorem}{Theorem}[section]

\theoremstyle{definition}

\theoremstyle{remark}

\begin{document}

%\jvol{00} \jnum{00} \jyear{2015} \jmonth{January}

\articletype{Review Article}

\title{Multi-parameter Quantum Metrology}

\author{
\name{Magdalena Szczykulska\textsuperscript{a}, Tillmann Baumgratz\textsuperscript{b}, and Animesh Datta\textsuperscript{b}$^{\ast}$\thanks{$^\ast$Corresponding author: animesh.datta@warwick.ac.uk},}
\affil{\textsuperscript{a}Department of Physics, University of Oxford, Oxford OX1 3PU, United Kingdom;\\
\textsuperscript{b}Department of Physics, University of Warwick, Coventry CV4 7AL, United Kingdom}
\received{v0.1 released December 2015}
}

\maketitle

\begin{abstract}
The simultaneous quantum estimation of multiple parameters can provide a better precision than estimating them individually. This is an effect that is impossible classically. We review the rich background of quantum-limited local estimation theory of multiple parameters that underlies these advances. We discuss some of the main results in the field and its recent progress. We close by highlighting future challenges and open questions.

%I am the abstract. \textbf{Length of abstract about 200 words!}
\end{abstract}

\begin{classcode}06.20.Dk Measurement and error theory, 03.65.Ta Foundations of quantum mechanics; measurement theory, 03.67.-a Quantum information, 42.50.-p Quantum optics
 %\textbf{(some more? \ldots please provide three to six PACS codes and their associated text)}
 \end{classcode}

\begin{keywords}
Quantum metrology; multi-parameter estimation; Cram{\'e}r-Rao bound; (quantum) Fisher information matrix; Heisenberg scaling 
%\textbf{(Please provide three to six keywords taken from terms used in your manuscript)}
\end{keywords}

{\abstractfont\centerline{\bfseries Index to information contained in this article}\vspace{12pt}
\hbox to \textwidth{\hsize\textwidth\vbox{\hsize 22pc
\hspace*{-12pt} {1.}    Why quantum metrology ?\\
\hspace*{7pt} {1.1.}   Why multiple parameters ?\\
{2.}    Multi-parameter estimation\\
\hspace*{10pt}{2.1.}  Zeitgeist \\
{3.}    Multi-parameter quantum metrology\\
\hspace*{10pt}{3.1.}  Unitary parameters \\
\hspace*{12pt} {3.1.1.}Recent advances \\
\hspace*{10pt}{3.2.}  Non-unitary parameters \\
\hspace*{10pt}{3.3.}  Saturating the multi-parameter QCRB \\
{4.}    Conclusion and outlook \\

}}}

%\textbf{Please note that the table of contents following the abstract in this guide is provided for information only. The typesetter will supply a Contents list in the style of the journal with the proof.}

\section{Why quantum metrology ?}

Metrology, as the science of measurements, has had an immense impact on the world we live in today. It has improved the quality of peoples' lives by enabling advances in areas such as navigation, telecommunication, transport and medicine~\cite{Quinn:2005aa} as well as facilitating trade, commerce and even high finance. It encapsulates a wide range of aspects, from defining the units of measurement and realising them in practice, to understanding phenomena and the fundamental limits that can be achieved in the precise estimation of parameters. These fundamental limits are set by the underlying theory of Nature -- quantum mechanics and therefore provide deep insights into the theory of quantum mechanics and hence Nature itself.

Metrology is thus the science -- and art -- of devising schemes that extract as precise as possible an estimate of the parameters associated with a system. A typical estimation process can be divided into three stages: probe preparation, interaction with the system and probe readout. For a given interaction with the system, the choice of probe states and measurements determines the precision with which one can measure the parameters of interest.  Appropriately chosen probe states ensure that the maximum amount of information about the parameters is encoded onto the probe, and appropriately chosen measurements maximise the amount of information that can be then extracted from the probe after acquiring this information. Even the most astutely designed and meticulously implemented scheme however, is affected by errors in the estimation process. The errors can either be systematic or statistical. Statistical errors of a stochastic nature can be reduced through repeated interactions between the probe and the system (corresponding to $M$ independent measurements), resulting in the typical statistical error (in standard deviation) scaling of $M^{-\frac{1}{2}}$. The origins of this scaling lies in the central limit theorem from probability theory, and is possible classically without the invocation of quantum mechanics. Given a probe of size (such as the number of particles or modes, energy) $N,$ the best classical possible scaling is the so-called standard quantum limit (SQL)\footnote{Although the SQL and the central limit theorem have the same quadratic dependence, they are of entirely different origins.}, whose error also scales as $N^{-\frac{1}{2}}$~\cite{Caves:1980aa}.

Once the stochastic noise is suppressed, quantum mechanics is the ultimate -- and most fundamental -- barrier to the precision of an estimation scheme. This inevitable limit is set by the quantum vacuum fluctuations and can only be overcome by invoking uniquely quantum mechanical techniques.  Quantum probes endowed with such non-classical correlations can attain the so-called Heisenberg limit, identified by a  $N^{-1}$ scaling in the standard deviation~\cite{Braunstein:1992aa,BRAUNSTEIN:1994aa}. This enhanced scaling, leading to a more precise estimation, is at the root of the appeal of quantum metrology. Quantum metrology can find application in scientific areas from astronomy -- detection of gravitational waves, to biology -- imaging of biological samples sensitive to the total illumination~\cite{Caves:1981aa,Taylor:2014aa}. It could be relevant for magnetic, electric, and gravitational field sensing, and more precise clock synchronisation protocols~\cite{Preskill2000,Chuang2000,Giovannetti:2004aa}.

Quantum metrology thus seeks scenarios where non-classical resources can provide improvements in the parameter estimation over the classical strategies and tries to identify the measurements that achieve quantum enhanced precisions. It must be understood that the quantum improvement can be availed only after all classical sources of stochastic noise have been suppressed. The most prominent example of this endeavour is the quest for the detection of gravitational waves using laser interferometry~\cite{Adhikari2014,Abbott2016}.

\subsection{Classical metrology}

Estimation theories can be categorised into global and local theories. In the global case, the parameter can be completely unknown and the estimation protocol enables finding the parameter of interest to some precision. Schemes based on the Bayesian theory, where the parameter is a random variable distributed according to some prior probability distribution, can be considered global~\cite{Demkowicz-Dobrzanski:2015aa,Toth:2014aa}. Examples of Bayesian precision bounds include the Ziv-Zakai and Weiss-Weinstein bounds~\cite{Ziv1969,Bell1997,Weinstein1988}. On the other hand, in some circumstances, we may already have a good knowledge of the interval where the true parameter value lies. In such cases, local approaches could be beneficial to further improve the precision and accuracy of the parameters of interest. Examples of local precision bounds include the Barankin~\cite{Barankin1949}, Holevo~\cite{Holevo:1982aa} and Cram{\'e}r-Rao bound. The last mentioned bound is the main topic of this article.

A statistical quantity capturing the performance of an estimation process is the variance of the estimator. A crucial result from probability theory, the Cram{\'e}r-Rao bound (CRB), states that the variance of an (unbiased) estimator is lower bounded by the inverse of the Fisher information (FI). The FI is a function of the probability distribution obtained at the end of an estimation process, and is of independent interest in probability theory, information theory, and information geometry~\cite{Amari2000}. More precisely, it is a distinguishability metric which provides a statistical distance on the space of probability distributions. It tells us how easily we can distinguish neighbouring probability distributions when separated by an infinitesimally small amount of the parameter value characterising the distributions. Therefore, the FI\footnote{The FI means different things in different contexts and the relationship between the FI and the entropy is well understood in the classical case. While the entropy is related to the volume of the typical set, the FI is related to the surface area of the typical set~\cite{Cover2006}.} captures the amount of ``information" about a given parameter in a probability distribution.

Estimators saturating the CRB are referred to as efficient. One of the difficulties of saturating the CRB is related to finding such efficient estimator whose existence is not guaranteed as explained in Theorem~(\ref{th:CRBsaturation}) at the beginning of Sec. (\ref{sec:2}), where we also discuss its asymptotic saturability. In principle, single as well as multi-parameter CRB can always be saturated.

\subsection{Back to quantum metrology}
In the quantum setting of the problem, the probability distribution depends on the input probe state (described by a density matrix) and the measurement (described by the positive operator valued measure). In this framework, maximising the FI over all valid quantum measurements leads to the quantum Fisher information (QFI) -- a distinguishability metric on the space of quantum states which quantifies the maximum amount of ``information" about a parameter attainable by a given probe state~\cite{Holevo:1982aa,Helstrom:1976aa,BRAUNSTEIN:1994aa}. Lower bounding the variance of an unbiased estimator with the inverse of the QFI is the so called quantum Cram{\'e}r-Rao bound (QCRB). Once the assumptions underlying local estimation theory are clearly stated, the QCRB provides the first step in understanding the fundamental limits of quantum enhanced metrology.

Single parameter QCRB can in principle be always saturated. However, an additional saturability problem arises in the quantum multi-parameter estimation due to the possible non-commutativity of quantum measurements. This additional aspect of quantum multi-parameter estimation is what makes quantum metrology interesting. The lower bound on the precision set by the QCRB can not be always attained when we try to simultaneously gain knowledge about multiple parameters. This is discussed further in Sec. (\ref{sec:sat}).

To illustrate the principle of quantum enhanced metrology, we consider single phase estimation in a noise-free environment using N00N states. For phase estimation protocols in the QCRB setting, states with high photon number variance are preferred since they maximise the QFI\footnote{QFI for a unitary parameter is proportional to the variance of the generator $\hat{G}$ and is given by $4M\left\langle\Delta{\hat{G}}^2\right\rangle$, where $M$ is the number of independent measurements. For the phase shift parameter, $\hat{G}$ is the number operator $\hat{n}$ and therefore as far as the QFI is concerned states with high photon number variance are preferred for phase estimation protocols. It is worth noting that the definition of the QFI assumes that the input state has a bounded variance and therefore it is not applicable for all probe states for instance $|\psi\rangle =\frac{\sqrt{3}}{2}\sum_m 2^{-m} |2^m\rangle$. In such circumastances, other bounds, such as the Ziv-Zakai bounds, are more suitable and find applications in particular for waveform estimation~\cite{Tsang:2014aa}.}. N00N states, $|\psi\rangle=(|N0\rangle+|0N\rangle)/\sqrt2$, exhibit this property with photon number variance of $N^2/4$ which gives the QFI of $MN^2$ and QCRB of $1/MN^2$ ($M$ is the number of independent measurements), and therefore they attain the desirable Heisenberg scaling. It is important to emphasise that N00N states achieve the QCRB only when one considers unbiased estimators and in particular if sufficient prior knowledge about the parameter value is available. The latter is due to the fact that this state is periodic in phase with period of $2\pi/N$ and therefore the phase needs to be known within this interval~\cite{Hall2012}. Further, N00N states are very fragile to loss. Loss of a photon in any of the two superposition elements quickly collapses the state onto the remaining element which completely destroys the superposition state. In this sense, states with more superposition elements, although smaller photon number variance, such as the Holland Burnett states~\cite{Holland1993} are more suitable. Additionally, N00N states are very hard to prepare and all the experimental realizations require post selection for $N>1$. Another class of states important for quantum metrology, which are more feasible to prepare experimentally, are squeezed coherent states. Squeezed coherent states exhibit reduced variance in one of the field quadratures in the expense of increased variance in the other quadrature so that to satisfy the Heisenberg uncertainty relation. This also offers enhancements in phase estimation protocols and one of the important applications includes detection of gravitational waves~\cite{Blair2016}. The topic of quantum enhanced estimation of a single parameter, typically phase, has been studied in great detail, and we direct the reader to the several extensive and excellent reviews~\cite{Giovannetti:2004aa,Giovannetti:2011aa,Demkowicz-Dobrzanski:2015aa,Toth:2014aa, Paris2009}. 

Before we progress on to the topic of multiple parameters, we must note that the promise of quantum enhancements in precision metrology is limited by the presence of noise, such as dephasing and dissipation in any experiment. Loss is an important limiting factor in photonic experiments, whereas phase diffusion typically plays a crucial role in experiments involving spins in atoms, ions, and vacancy centres. Although it is known that the Heisenberg scaling would eventually vanish in the presence of noise to match the SQL~\cite{Dobrzanski2013}, it is still possible to gain advantage over the classical schemes. The attainable precision in such realistic cases is an area of active investigation and some progress has been made in obtaining general upper bounds for the QFI corresponding to a single phase estimation in the presence of noise~\cite{Escher:2011aa,Escher:2012aa, Demkowicz-Dobrzanski2012, Huelga1997, Matsuzaki2011, Macieszczak2015, Smirne2016}.

\subsection{Why multiple parameters?}

For equivalent resources, simultaneous quantum estimation of multiple phases or in general, parameters corresponding to non-commuting unitary generators, provides better precision than estimating them individually~\cite{Humphreys:2013aa, Baumgratz:2015aa}. This has generated interest in multi-parameter quantum metrology in a variety of scenarios and contexts~\cite{Spagnolo:2012aa,Vaneph:2013aa,Yue:2014aa,Zhang:2014aa,Gao:2014aa,Tsang:2014aa,Zhang:2014ab,Yao:2014aa,Zuppardo:2015aa,Kok:2015aa,Klimov2005}. However, the myriad motivations for studying multi-parameter quantum metrology are deeply interleaved and intertwined. Nevertheless, the following presents a broad delineation of at least three broad seams of interest:

\begin{enumerate}
\item \textit{Physics}: The measurements that maximise the QFI corresponding to multiple parameters need not necessarily commute. Thus, the enhancements in precision metrology promised by quantum mechanics might eventually be thwarted by quantum mechanics. It is of principal interest in quantum information theory as it explores the information extracting capabilities of quantum measurements, and provides a rich new testing bed for understanding the nature of quantum measurements in great generality. High precision estimation is also beginning to play a role in the detection of novel phenomena such as gravitational waves~\cite{Abbott2016} and should lead to discoveries yet unknown in other areas of fundamental physics. 

\item \textit{Mathematics}: The quantum Fisher information matrix -- the multi-parameter extension of the QFI -- is a `metric' on the space of quantum states. Although it is not unique as in the classical case, it is minimal among all monotone metrics~\cite{Petz:1996aa}. This makes it not only a quantity of inherent interest in quantum metrology, but also capable of unlocking novel features of the space of quantum states whose study underlies all of quantum information theory, non-commutative geometry and quantum information geometry~\cite{Ercolessi2012, Ercolessi2013, Contreras2016}. 

\item \textit{Technology}: Numerous high level applications intrinsically involve multiple parameters. Quantum enhanced schemes for imaging~\cite{Tsang2016}, microscopy, spectroscopy to high precision sensors for classical electric, magnetic, gravitational fields cannot be developed without a clear understanding of multi-parameter quantum metrology. Eventually, highly precise characterisation of components for fault-tolerant quantum technologies~\cite{Martinis:2015aa} and quantum information science~\cite{Childs:2000aa} might also benefit from multi-parameter quantum metrology.

\end{enumerate}
To make some of these motivations more concrete and delve into the status of the field in greater details, we first describe the problem mathematically. As we point out along the way, the inception of multi-parameter quantum metrology as a field is just as rich as its future prospects.

\section{\label{sec:2}Multi-parameter estimation}
%General, brief, overview.

The central task here is of estimating a set of parameters $\bm{\theta}=(\theta_1,\cdots,\theta_d)^T \in \mathbb{R}^d.$ The precision of any estimator $\bm{\hat{\theta}}$ of $\bm{\theta}$ is given by the mean square error $\mathbb{E}\left[(\bm{\theta}-\bm{\hat{\theta}})(\bm{\theta}-\bm{\hat{\theta}})^T\right],$  the expectation value of squared difference. For unbiased estimators, the mean squared error is equal to the covariance matrix $\cv(\bm{\hat{\theta}}).$ One of the central results of the classical probability theory, the Cram\'er-Rao inequality, places a lower bound on the covariance matrix
\be
\cv(\bm{\hat{\theta}}) \geq C\left(\bm\theta\right)^{-1}
\ee 
where $C\left(\bm\theta\right)$ is the Fisher information matrix with elements 
\be
\label{eq:FI}
\left[C\left(\bm\theta\right)\right]_{ij}=\mathbb{E}\left[\left(\frac{\partial}{\partial \theta_i}\log p(x,\bm{\theta})\right)\left(\frac{\partial}{\partial \theta_j}\log p(x,\bm{\theta})\right)^T\right],
\ee
 which depend on the probability distribution $p(x,\bm{\theta})$ of the outcomes $x$. The Cram\'er-Rao inequality applies only to well behaved probability distributions which satisfy the following regularity condition,

\begin{equation}
\label{eq:RegularityCondition}
\mathbb{E}\left[\frac{\partial}{\partial \theta_i}{\log{ p(x,\bm{\theta})}}\right]=0\hspace{10mm}\text{for}\hspace{2mm}\text{all}\hspace{2mm}\bm{\theta},
\end{equation}
 
where the expectation value is taken with respect to $p(x,\bm\theta)$. Additionally, the estimator $\bm{\hat\theta}$ saturating the CRB is locally unbiased and therefore must satisfy
\begin{equation}
\label{eq:UnbiasedEstimator}
\mathbb{E}\left[\bm{\hat{\theta}}\right]=\bm\theta
\end{equation}
in the neighbourhood of the true value of the parameters\footnote{~A CRB with biased estimators can also be defined \cite{Cover2006}.}.
The CRB is proven using the Cauchy-Schwarz inequality~\cite{Helstrom:1976aa}. The condition for existence of a locally unbiased estimator saturating the bound is stated in Theorem~(\ref{th:CRBsaturation}). If such a locally unbiased estimator exists, the bound can always be saturated using the maximum likelihood estimator. Although this saturation by maximum likelihood estimator is in principle asymptotic in the number of experiments, it has found widespread practical use since a finite number of data points usually provides satisfactory performance~\cite{Braunstein1992}. However, identifying the conditions for the saturation of the CRB is an intricate topic with a variety of technicalities. Most deal with the differentiability of the probability distribution function $p(x,\bm{\theta}),$ the most common being the notion of the differentiability of the quadratic mean~\cite{LeCamYang2000}. The issue of finding the estimator saturating the CRB and the associated saturability is not of quantum origin and cannot be resolved using quantum mechanics.

%This is a classical problem which has no relations to quantum mechanics. It is also important to stress that the unbiasedness restriction only applies locally and therefore the estimation strategy needs to hold true only in the vicinity of the true parameter value. As a result, we often need to have a good knowledge of the true parameter itself. Although the unbiasedness condition can be very restrictive, it can be shown that the maximum likelihood estimator is asymptotically efficient, given that such estimator exists for the considered probability density function (see Eqs.~(\ref{eq:RegularityCondition}), ~(\ref{eq:UnbiasedEstimator}) and ~(\ref{eq:SaturationCondition})). How large sample size $M$ is required for the assymptotic efficiency to be applicable depends on the considered scenario, however, finite $M$ is usually sufficient~\cite{Braunstein1992,Geyer2012}.
 
\begin{theorem}
\label{th:CRBsaturation}
Given that the probability distribution satisfies Eq. (\ref{eq:RegularityCondition}), a local unbiased estimator $\bm{\hat\theta}$ saturating the CRB exists iff~\cite{Kay1993},
\begin{equation}
\label{eq:SaturationCondition}
\frac{\partial}{\partial\bm\theta}\log{ p(x,\bm{\theta})}=C\left(\bm\theta\right)\left({\bm{\hat{\theta}}}-\bm{\theta}\right).
\end{equation}
\end{theorem}

The quantum version of estimation theory begins with a quantum state $\rho_0$ which undergoes an evolution depending on $\bm{\theta}$. The resulting state $\rho_{\bm{\theta}}$ is measured by a set of positive operator valued measures (POVMs) $\{\Pi_x\},$ leading to probabilities given by $p(x,\bm{\theta}) = \tr{\rho_{\bm{\theta}}\Pi_x}.$ All the information about the parameters $\bm{\theta}$ is now encapsulated in the probability distribution $p(x,\bm{\theta}),$ and can be used to estimate the parameters with a precision given by the classical Fisher information in Eq.~(\ref{eq:FI}). However, the Fisher information now also depends on the POVMs $\{\Pi_x\},$ which stands in the way of obtaining a fundamental quantum limit to the covariance $\cv(\bm{\theta}).$ The mathematical task is thus that of maximising the classical Fisher information over all POVMs giving rise to the QFI, and the conceptual challenge lies in extending the notion of a derivative to the space of quantum states. 

\subsection{Zeitgeist }
\label{sec:zeit}

In 1967, Helstrom defined a family of operators  $L_i$ called symmetric logarithmic derivatives (SLDs) to capture the notion of the differential of a quantum state as
\be
 L_i \rho_{\bm{\theta}} + \rho_{\bm{\theta}} L_i = 2 \frac{\partial \rho_{\bm{\theta}}}{\partial \theta_i}
\ee
leading to the multi-parameter quantum Cram\'er-Rao bound (QCRB)~\cite{Helstrom:1967aa,Helstrom:1968aa}
\be
\label{eq:QCRB}
\cv(\bm{\theta}) \geq Q^{-1},
\ee
where $ Q_{ij} = \tr{\rho_{\bm{\theta}}(L_iL_j + L_jL_i)}/2$ is called the quantum Fisher information matrix (QFIM)\footnote{This matrix is sometimes referred to as the (symmetric) quantum Fisher information, Helstrom information, Helstrom matrix.}. He showed that the individual parameter $\theta_i$ can be estimated with a variance lower bounded by the inverse of $\tr{\rho_{\bm{\theta}} L_i^2},$ and a POVM attaining this precision is given by the eigenvectors of the SLD $L_i.$ He did not consider the collective saturation of the bound for all the parameters simultaneously, but identified the central problem in multi-parameter quantum estimation theory -- that the optimal POVMs corresponding to different $L_i$ need not necessarily commute. 

In 1972, Belavkin in the Soviet Union first exploited the Cram\'er-Rao bound to formulate generalised Heisenberg uncertainty principle for quantities such as time and energy~\cite{Belavkin:1976aa}, extending the early work of Mandelstam and Tamm~\cite{Mandelstam1945}. To that end, he defined the right logarithmic derivative (RLD) $R_i$ as~\cite{Belavkin:1976aa}
\be
\rho_{\bm{\theta}} R_i = \frac{\partial \rho_{\bm{\theta}}}{\partial \theta_i}.
\ee
In 1973, Yuen and Lax defined the same quantity to study the attainment of the multi-parameter QCRB with the family of coherent states in thermal noise~\cite{YUEN:1973aa}. They showed that  to saturate the multi-parameter quantum Cram\'er-Rao inequality, it may sometimes be necessary to include the possibility of non-Hermitian operators (this is done by considering measurements on a larger system). This is a consequence of the fact that while the SLDs are guaranteed to be Hermitian, the RLD need not be so. Allowing for a non-Hermitian RLD, the estimation of two real parameters can be recast as the estimation of one complex parameter. Furthermore, the Cram\'er-Rao inequality based on the RLD may not be attainable, even in the single-parameter scenario as the optimal measurements may not be measurable. In 1974, Helstorm and Kennedy studied non-commuting observables in the multi-parameter setting and developed the notion of the most informative bound~\cite{HELSTROM:1974aa}. Holevo later expanded the results of Belavkin, and Yuen and Lax to the estimation problem of the expectation parameter of family of quantum Gaussian states~\cite{Holevo:1982aa}, including those involving the RLD. He also obtained a lower bound on the mean square error which can be applied in great generality, and is now called the Holevo bound~\cite{Holevo:1982aa}.

Also in 1968, Braginskii realized that the expected amplitude of gravitational wave-induced oscillations of a bar detector signal mode would be on the order of the zero point oscillations of this mode, as predicted by quantum mechanics. Thus, in order to observe gravitational waves, one has to treat a detector quantum-mechanically and consequently, there will be a quantum back action, setting a limitation on the achievable sensitivity, the SQL~\cite{Braginskii1968}. By the late 1970s, Braginskii and coworkers were seeking different detectors such as ground-based optical interferometers to circumvent the SQL in gravitational wave detectors~\cite{Braginsky1978}. By 1980, Caves had shown that the limits to the precision of phase estimation in an interferometer is set by the vacuum fluctuations entering its empty port~\cite{Caves1980}. 

Helstrom, Kennedy, Lax and Yuen had been interested in the limits of optical communication engineering and radar systems. The eventual quantum nature of the electromagnetic radiation had led them to quantum estimation theory. Belavkin and Holevo were largely driven towards a deeper understanding of quantum mechanics and quantum measurements. The designers of gravitational wave detectors, and later interferometers in the late 1970s and early 1980s headed into quantum estimation theory to better understand the ultimate limits of their instruments. In Japan, information geometry was being developed in the 1980s. Information geometry applies the methods of differential geometry to probability theory by considering probability distributions as points on a Riemannian manifold, with the Riemannian metric being given by the Fisher information in Eq.~(\ref{eq:FI})~\cite{Amari2000}. And the methods were ready to attack the probability distributions arising from quantum systems by the middle of the 1990s. 

Multi-parameter quantum metrology has thus emerged as the conflux of several disparate streams of scientific objectives and aspirations. 

\section{Multi-parameter quantum metrology}

In 1994, Braunstein and Caves brought the methods of quantum estimation theory for a single parameter to quantum physics~\cite{Braunstein:1992aa,BRAUNSTEIN:1994aa}. The main contribution of this work lies in the separation of the classical and quantum optimisation necessary for the saturation of the quantum Cram\'er-Rao bound.  In the same year, Fujiwara addressed also the same problem, but limited to pure states~\cite{Fujiwara:1994ab,Fujiwara:1995aa}. He also addressed the theory of multi-parameter estimation for pure states based on the RLD~\cite{FUJIWARA:1994aa,Fujiwara:1999aa}. In 1995, Massar and Popescu constructed an optimal measurement to determine two parameters that identify a specific pure state, and also showed that the optimal measurement is an entangled measurement over all $N$ probes~\cite{Massar:1995aa} in an answer to a question by Peres and Wootters~\cite{Peres:1991aa}. 

In 1996, the mathematical agitation between the SLD and RLD bounds that afflicted the multi-parameter quantum metrology was dramatically resolved by Petz and coworkers~\cite{Petz1996,Petz:1996aa}.  They show that all stochastically monotone Riemannian metrics are characterized by means of operator monotone functions and prove that there exist a maximal and a minimal among them. In particular, the minimal one is none other than that given by the QFIM. Invoking methods from operator theory, these results endowed the QFIM with a new fundamental character. In 1997, another important result was obtained -- Matsumoto showed that a multi-parameter quantum Cram\'er-Rao inequality can always be saturated for pure states~\cite{Matsumoto:1997aa,Matsumoto:2002aa}\footnote{Parts of the Matsumoto's thesis appear in chapter 20 of Ref.~\cite{Hayashi:2005aa}.}.

A natural context in which multi-parameter quantum metrology has an operational interpretation is quantum state estimation. Since the multi-parameter QCRB is attainable only when the SLDs commute in the expectation value (See Sec.~\ref{sec:sat}), Holevo~\cite{Holevo:1982aa} obtained a lower bound on the MSE but it remained an open problem whether this bound is achievable for mixed states. In the last few years, Gu\c{t}\u{a} and coworkers have developed the quantum local asymptotic normality theory for quantum state estimation, and one of its consequences (up to some technicalities) is the achievability of the Holevo bound~\cite{Kahn2008,Guta2008,Kahn2009}. 

\subsection{Unitary parameters}

The estimation of a single phase has always been the most ubiquitous form of the problem from a physicists' perspective. This has been driven by the central role that interferometry, which measures a relative phase, plays in numerous areas of physics, and spurred on by the impetus to improve the sensitivity of gravitational wave interferometers~\cite{Danilishin2012}. More generally,  estimating a unitary operator with fidelity as the figure of merit has been studied~\cite{Acin:2001aa}. Similarly, simultaneous estimation of multiple phases was considered in~\cite{Macchiavello2003}. Further, the strategy requires a reference system with entanglement between the system and a reference system. This highlights an important issue in the field -- that for different cost functions different measurements will be optimal~\cite{Barndorff-Nielsen:2000aa}. This work also discussed the non-attainability of the multi-parameter Cram\'er-Rao bound because the optimal measurements might not commute. In the estimation of commuting unitary operators $U,$ Ballester showed that no advantage is afforded by using entangled input states~\cite{Ballester:2004ab}. Note that in this setup, the quantum system was divided into two parts, one of which sensed the unitary, while the other half remained untouched. In the same setup, entangled measurements enabled an improvement of the precision by a factor $2(d+1)/d$, where $d$ is the dimension of the Hilbert space on which the unitary acts~\cite{Ballester:2004aa}. For non-commuting unitaries, the transmission of a reference frame through a quantum channel made out of $N$ spins has been studied as a $SU(2)$ estimation problem, leading to an average error that obeys a Heisenberg scaling \cite{Bagan:2001aa,Fujiwara:2001ab,Kolenderski08}. 

The $SU(2)$ estimation problem has also been studied using methods from group theory such as equivalent representations and multiplicity spaces, showing the requirement of entanglement between spaces, where the action of the group is irreducible and spaces, where the action of the group is trivial~\cite{Chiribella:2004aa}.  A similar result holds for the optimal Bayesian estimation of an unknown transformation with a quantum-enhanced Heisenberg scaling~\cite{Chiribella:2005aa}.  $SU(d)$ estimation has also been studied, but the $d$ dependence of the variance was not explored~\cite{IMAI:2007aa,Kahn:2007aa}. In \cite{Chiribella:2006aa}, the authors discussed the joint estimation of real squeezing and displacement in phase space. They found optimal measurements for a joint estimation that maximise the likelihood function.  They also highlighted the nonunimodularity of the group as playing a vital role in the estimation process, and once again noted the value of quantum entanglement in precision estimation. The same was noted in the estimation of displacements, a complex parameter, in phase space~\cite{Genoni:2013aa}. A recent experiment demonstrating a quantum-enhanced tomography of an unknown unitary process is outlined in~\cite{Zhou2015}.

\begin{wrapfigure}{c}{25mm}
%  \centering
  \vspace{-1.5cm}
    \includegraphics[scale=0.2,viewport= 0 0 400 600,clip]{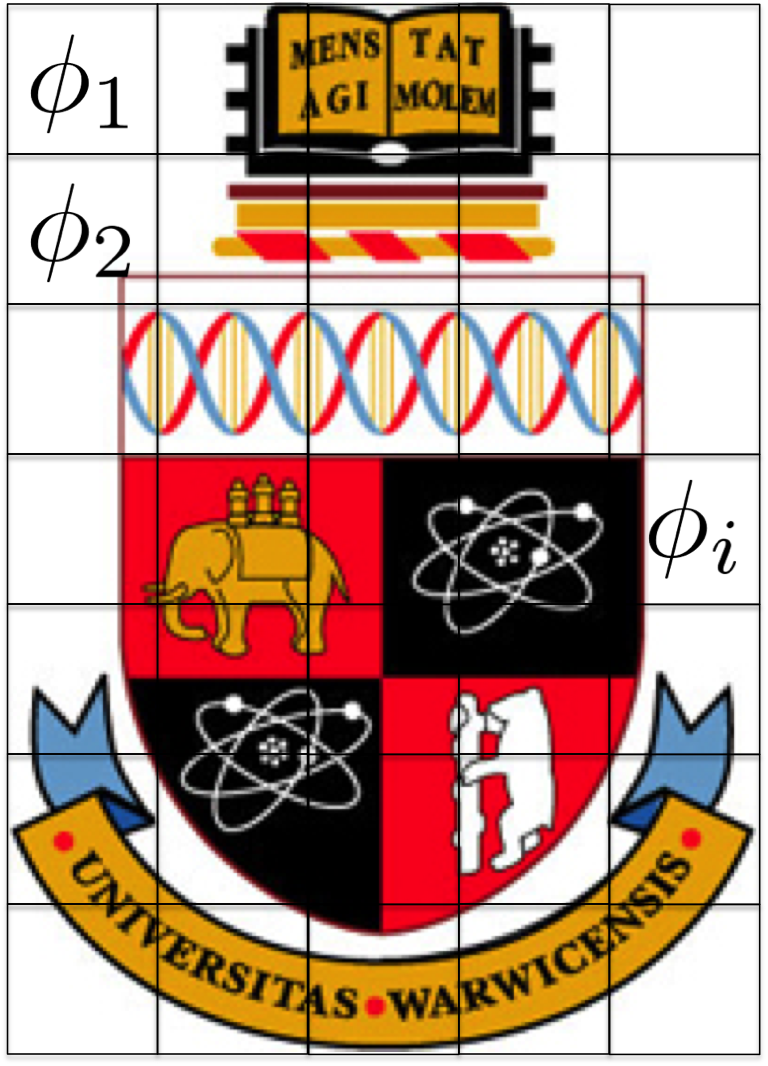}
  \caption{Phases $\phi_i$ label the pixels in an image}
  \label{fig:pixels}
\end{wrapfigure}

\subsubsection{Recent advances}

In 2013, Humphreys and co-workers showed that for a fixed number of photons, the precision in estimating a certain number of phases across independent modes is better if they are estimated simultaneously rather than individually~\cite{Humphreys:2013aa}. They also showed that the total variance decreases linearly with the number of parameters. This makes multi-parameter very attractive from a technological point of view. Of course, the investment needed to harness this advantage is the generation of entangled quantum states of an increasing number of modes. However, this could be worthwhile in imaging applications where an object can be considered as a collection of independent pixels as shown in Fig.~(\ref{fig:pixels}). Experimental efforts have been made to estimate phases in multi-arm interferometers as first steps towards such a realisation~\cite{Spagnolo:2012aa,Ciampini15}. The initial proposal has also been studied in realistic circumstances and although the enhancement of multi-phase estimation eventually reduces to the SQL in the presence of loss, an advantage still remains if the loss figure is not too high and robust states are employed~\cite{Yue:2014aa}.

A similar `multi-parameter' advantage, proportional to the number of parameters, was shown in Ref.~\cite{Baumgratz:2015aa} in the estimation of fields in 3 dimensions. The technique presented applies to any number of dimensions and works in spite of the non-commutativity of the generators. This covers scenarios of interest such as magnetic, electric, or gravitational field imaging in 3 dimensions simultaneously. It is mathematically identical to the estimation of hamiltonians as in Ref.~\cite{Skotiniotis:2015aa}, although this work does not exploit its multi-parameter aspects. This aspect was studied in the physical context of a non-demolition measurement of a Bose-Einstein condensate in a double-well optical cavity~\cite{Zuppardo:2015aa}. The work in~\cite{Liu2016} also investigates the multi-parameter aspect of phase estimation, but with entangled coherent states. It finds that the simultaneous estimation can provide a better precision than the independent estimation and that the entangled coherent states outperform the generalised N00N states in the equivalent estimation scenario. A case of quantum-enhanced multi-phase estimation using Gaussian inputs has been studied in~\cite{Gagatsos2016}. The work shows that assuming equally squeezed input states and an orthogonal interferometer, the simultaneous phase estimation strategy is always better than the individual phase estimation with the figure of merit being trace of the QFI matrix.

\subsection{Non-unitary parameters}

While pure states and unitary transformations have occupied most of the attention in the realm of quantum metrology, the full characterisation of a system would also require the estimation of decoherence parameters. Simultaneous estimation of all the parameters yield a better understanding of the underlying system, and include parameters such as diffusion and loss. This is an improvement on the typical strategy of estimating the decoherence in independent experiments and using that value to optimise phase estimation in the presence of decoherence~\cite{Dorner2009,Knysh2011,Knysh:2013aa}. 

The estimation of decoherence parameters is intrinsically linked to mixed states since they are the end result of a decoherent evolution. The quantum estimation theory of mixed states has been covered in the early work of Helstrom and others. However, their emphasis on coherent states avoided explicit investigation of mixed states. Braunstein and Caves also made  explicit the distinction between pure and mixed states. One of the early works within the information geometry framework was by Fujiwara~\cite{Fujiwara:1995aa,Fujiwara:1999aa}. Optimal estimation of qubit mixed states (all the components of a Bloch vector) was studied in Refs.~\cite{Fujiwara:2001aa,Fujiwara:2003aa,Bagan:2006aa,Hayashi:2008aa}, and  information geometry was employed to study the estimation of multiple parameters from Markovian dynamics~\cite{Guta16} and Gaussian states~\cite{Monras:2010aa,Monras:2011aa}. Gu\c{t}\u{a} and others employed local asymptotic normality for the estimation of mixed quantum states~\cite{Kahn2009}. 

One of the interesting aspects of multi-parameter estimation is the tradeoff in the attainable precisions that arises due to the possible non-commutativity of optimal measurements, for instance in the simultaneous estimation of phase and loss in optical interferometry~\cite{Crowley:2014aa}. Such a tradeoff can also arise if the dimension of the Hilbert space is not enough to accommodate all the parameters. This was studied by Gill and Massar in 2000 in the problem of estimating parameters related to quantum state estimation~\cite{Gill:2000aa}. This was later identified in the simultaneous estimation of phase and dephasing for qubits~\cite{Vidrighin:2014aa}. A specific class of measurements, called Fisher-symmetric informationally complete measurements, that can saturate these tradeoffs have also been studied recently~\cite{Li15}. 

\subsection{Saturating the multi-parameter QCRB}
\label{sec:sat}

The multi-parameter QCRB is an inequality, and identifying the conditions of its saturation is salient to its understanding. Since, as stated in Sec. (\ref{sec:zeit}), the SLDs corresponding to the different parameters need not commute, the saturation of the multi-parameter QCRB is not assured. This is an issue that does not arise in single parameter estimation theory, where saturation is guaranteed. For multiple parameters, a sufficient condition for the saturation is $[L_i,L_j]=0,$ the commutation of the SLDs. However, this is not the necessary condition and it is not obvious what this condition is in general. It is known that the Holevo bound can be asymptotically attained within the framework of local asymptotic normality (LAN), where the model converges to a Gaussian shift model~\cite{Gill2013}. The precision associated with the QCRB is always smaller or equal to the precision associated with the Holevo bound. When the Holevo bound coincides with the QCRB based on SLDs, the necessary and sufficient condition is
\be
\label{eq:lan}
 \tr{ \rho_{\bm{\theta}}[L_i, L_j]}=0.
\ee
In the framework of LAN, $N$ copies of the state $\rho_{\bm{\theta}}^{\otimes N}$ tend to a locally continuous variable system -- the product of a Gaussian probability density function and a tensor product of uncorrelated single mode quantum Gaussian states. The commutation relation of the collective modes of these latter Gaussian states is given by~\cite{Gill2013} $ \tr{ \rho_{\bm{\theta}}[X_i, X_j]}$, where $X_i$ is the collective variable. In the instances when the Holevo bound is the same as the SLD quantum Cram\'er-Rao bound this becomes $\tr{ \rho_{\bm{\theta}}[L_i, L_j]}$. The parameters of a single mode Gaussian state can be measured simultaneously if the commutation relation vanishes. This leads to the necessity of Eq.~(\ref{eq:lan}) for saturating the multi-parameter QCRB at least in these special circumstances. Note that the convergence to the saturation is asymptotic, and can be attained by the maximum likelihood estimator~\cite{Geyer2012}. To the knowledge of the authors, an exact and general relationship between the Holevo bound and the QCRB is not established. However, there has been studies connecting the Holevo bound to the QCRB in special cases. Ref.~\cite{Suzuki2015} investigates such connection for a two-parameter qubit model and gives a condition for when these two bounds are equivalent.

For quantum estimation using pure states, the multi-parameter QCRB can however be always saturated asymptotically~\cite{Matsumoto:2002aa}. The underlying necessary and sufficient condition is still that of commuting SLDs in the expectation value, and for a general hamiltonian estimation the optimal measurement can be constructed explicitly~\cite{Baumgratz:2015aa}.

\section{Conclusions and Outlook}

The task of quantum metrology is to obtain as precise as possible an estimate of a set of parameters using quantum probes. The choice of the measurement is a vital ingredient in this process. This is brought to the fore in multi-parameter quantum metrology. This is precisely why multi-parameter quantum metrology provides a fertile ground for understanding quantum measurements. 

The potential applications of multi-parameter quantum metrology are wide beyond its appeal as a domain for a deeper understanding of quantum mechanics. It has prospective appeal in the development of quantum technology itself. In a fault-tolerant quantum computer, the qubits and their logic interactions must have errors below a threshold of $10^{-18}.$ Characterising such a system will require multi-parameter quantum metrology at the level of 1- and 2-qubit gates~\cite{Martinis:2015aa}. Another area could be the understanding of the hamiltonians driving quantum phase transitions. Since these are zero temperature phenomena, their direct probing will necessarily require probes with minimal energy and disturbance, the kind provided by quantum metrology. Other scenarios for multi-parameter quantum metrology could include the imaging of electric, magnetic, gravitational and other fields in 3 or more dimensions, as well as multimode quantum imaging~\cite{Kolobov1999}. All of these have fundamental scientific as well as technological applications. 

The above applications, allied with the intrinsic richness and variety of the topics that touch upon multi-parameter quantum metrology, make it a topic worth pursuing. Open questions abound. One of the first should be a systematic study of multi-parameter quantum metrology in the mould of Ref.~\cite{Giovannetti:2006aa}, where the authors discuss all four possible scenarios with respect to probe states and measurements -- classical and classical, classical and quantum, quantum and classical, and quantum and quantum. This should clarify the role of quantum correlations in circumventing the tradeoffs that arise in multi-parameter quantum metrology. Another open question is the relation of the Holevo bound to the Cram\'er-Rao bound, and there have been some recent results for special cases such as the two-parameter qubit estimation problem~\cite{Suzuki2015}. One very fruitful area could be the use of information geometry methods to identify the tradeoff relations in multi-parameter quantum metrology and optimal measurements necessary to saturate them. A final open challenge could be to go possibly beyond the differential approach of Helstrom and information geometry, and identify measurements that minimise the mean square error over all the parameters.

\section*{Acknowledgements}

AD thanks Mihai Vidrighin for bringing Ref.~\cite{Gill:2000aa} to his attention and M\u{a}d\u{a}lin Gu\c{t}\u{a}, Rafa\l{} Demkowicz-Dobrza\'{n}ski and Michael Hall for several interesting discussions. This work was supported, in part, by the UK EPSRC (EP/K04057X/2), the National Quantum Technologies Programme (EP/M01326X/1, EP/M013243/1), and the DSTL (DSTLX-1000091903).

%\section{Some References in detail}
%%\cite{YUEN:1973aa,HELSTROM:1974aa,HAYASHI:2006aa,FUJIWARA:1994aa,Fujiwara:2001ab,IMAI:2007aa,Fujiwara:2003aa,Fujiwara:2001aa,Humphreys:2013aa,Crowley:2014aa,Vidrighin:2014aa,Gill:2000aa,BRAUNSTEIN:1994aa,Giovannetti:2011aa,PARIS:2009aa,Hayashi:2008aa,Monras:2010aa,Toth:2014aa,Demkowicz-Dobrzanski:2015aa,Giovannetti:2006aa,Genoni:2013aa,Yue:2014aa,Zhang:2014aa,Gao:2014aa,Tsang:2014aa,Kok:2015aa,Giovannetti:2004aa,Helstrom:1976aa,Skotiniotis:2015aa,Matsumoto:2002aa,Chiribella:2005aa,Zhang:2014ab,Pfister:2015aa,Martinis:2015aa,Belavkin:1976aa,Chiribella:2006aa,Young:2009aa,Childs:2000aa,Watanabe:2010aa,Monras:2011aa,Vaneph:2013aa,Yao:2014aa,Zuppardo:2015aa,Gill:2013aa,Knysh:2013aa,Holevo:1982aa,Spagnolo:2012aa,Genoni:2008aa,Cramer:1999aa,Helstrom:1967aa,Zanardi:2008aa,Helstrom:1968aa,Personick:1970aa,Tsang:2011aa,Macchiavello:2003aa,Jones:1991aa,Jones:1994aa,Slater:2001aa,Fujiwara:1995aa,Fujiwara:1999aa,Petz:1996aa,Kahn:2007aa,Barndorff-Nielsen:2000aa,Ballester:2004aa,Massar:1995aa,Peres:1991aa,Chiribella:2004aa,Fujiwara:1994ab,Nagaoka:1989aa,Nagaoka:1989ab,Hirota:1997aa,Fujiwara:2005aa,Matsumoto:1997aa,Baumgratz:2015aa,Quinn:2005aa}

%\nocite{*} 

\bibliographystyle{tADP}
\bibliography{BibliographyReviewMultiMetrology}

%\appendices
%\section{This is the title of the first appendix}
%\section{This is the title of the second appendix}

\end{document}